\pdfoutput=1
\documentclass[twocolumn,preprintnumbers,amsmath,amssymb,letter]{revtex4}
\usepackage{graphicx}
\usepackage{dcolumn}

\newcommand{\ttt}{\texttt}

\begin{document}

\preprint{KRS-2009-01}

\title{A Graph Analysis of the Linked Data Cloud}

\author{Marko A. Rodriguez}
\affiliation{Semantic Network Research Group \\
		Knowledge Reef Systems Inc. \\
		Santa Fe, New Mexico 87501 }
		
\date{\today}

\begin{abstract}
The Linked Data community is focused on integrating Resource Description Framework (RDF) data sets into a single unified representation known as the Web of Data. The Web of Data can be traversed by both man and machine and shows promise as the \textit{de facto} standard for integrating data world wide much like the World Wide Web is the \textit{de facto} standard for integrating documents. On February 27$^\text{th}$ of 2009, an updated Linked Data cloud visualization was made publicly available. This visualization represents the various RDF data sets currently in the Linked Data cloud and their interlinking relationships. For the purposes of this article, this visual representation was manually transformed into a directed graph and analyzed.
\end{abstract}

\maketitle{}

\section{Introduction}

The World Wide Web is a distributed document and media repository \cite{lee94}. Hyper-Text Markup Language (HTML) documents reference other HTML documents and media (e.g.~images, audio, etc.) by means of an \ttt{href} citation. The resulting document citation graph has been the object of scholastic research \cite{bowtie:huberman1999,bowtie:broder} as well as a component utilized in web page ranking \cite{anatom:brin1998}. Similarly, the Semantic Web is a distributed resource identifier repository \cite{pubsem:lee2001}. The Resource Description Framework (RDF) serves as one of the primary standards of the Semantic Web \cite{rdfintro:miller1998}. RDF provides the means by which Uniform Resource Identifiers (URI) \cite{uri:berners2005} are interrelated to form a multi-relational or edge labeled graph. If $U$ is the set of all URIs, $L$ is the set of all literals, and $B$ is the set of all blank (or anonymous) nodes, the the Semantic Web RDF graph is defined as the set of triples
%%%
\begin{equation*}
	G \subseteq (U \cup B) \times U \times (U \cup L \cup B).
\end{equation*}
%%%
Given that the URI is the foundational standard of both the World Wide Web and the Semantic Web, the Semantic Web serves as an extension to the World Wide Web in that it provides a semantically-rich graph overlay for URIs. Thus, the Semantic Web moves the Web beyond the simplistic \ttt{href} citation into a rich relational structure that can be utilized for numerous end user applications.

The Linked Data community is actively focused on integrating RDF data sets into a single connected data set \cite{berners:ldata2006}. The Linked Data model allows
%%%
\begin{small}
\begin{quote}
``[any man or machine] to start with one data source and then move through a potentially endless Web of data sources connected by RDF links. Just as the traditional document Web can be crawled by following hypertext links, the Web of Data can be crawled by following RDF links. Working on the crawled data, search engines can provide sophisticated query capabilities, similar to those provided by conventional relational databases. Because the query results themselves are structured data, not just links to HTML pages, they can be immediately processed, thus enabling a new class of applications based on the Web of Data." \cite{linkeddata:bizer2008}
\end{quote}
\end{small}

While the Linked Data community has focused on providing a distributed data structure, they have not focused on providing a distributed process infrastructure  \cite{rodriguez:distributed2008}. Unfortunately, if only a data structure is provided, then processing that data structure will lead to what has occurred with the World Wide Web: a commercial industry focused on downloading, indexing, and providing search capabilities to that data. For the problem space of keyword search, this model suffices. However, the RDF data model is much richer than the World Wide Web citation data model. If data must be downloaded to a remote machine for processing, then only so much of the Web of Data can be processed in a reasonable amount of time. This ultimately limits the sophistication of the algorithms that can be executed on the Web of Data. The RDF data model is rich enough to conveniently support the representation of relational objects  \cite{activerdf:oren2008} and their computational instructions \cite{rodriguez:gpsemnet2007}. Moreover, with respect to searching, the RDF data model requires a new degree of sophistication in graph analysis algorithms \cite{semrank:boan2005}. For one, the typical PageRank centrality calculation is nearly meaningless on an edge labeled graph \cite{grammar:rodriguez2007}. To leave this algorithmic requirement to a small set of search engines will ultimately yield a limited set of algorithms and not a flourishing democracy of collaborative development. As a remedy to this situation, a distributed process infrastructure (analogous in many ways to the Grid \cite{grid:foster2004}) may be a necessary requirement to ensure the accelerated, grass roots use of the Web of Data, where processes are migrated to the data, not data to the processes. In such a model, computational clock cycles are as open as the data upon which they operate.

With respect to the Web of Data as a distributed RDF data structure, this article presents a graph analysis of the March 2009 Linked Data cloud visualization that was published on February 27, 2009 by Chris Bizer.\footnote{The March 2009 Linked Data cloud visualization is available at: \ttt{http://tinyurl.com/b4vfbq}.} The remainder of this article is organized as follows. \S \ref{sec:construction} articulates how the Linked Data cloud graph was constructed from the February 27$^\text{th}$ Linked Data cloud visualization. \S \ref{sec:statistics} provides a collection of standard graph statistics for the constructed Linked Data cloud graph. Finally \S \ref{sec:structural} provides a more in-depth analysis of the structural properties of the graph.

\section{Constructing the Linked Data Cloud Graph\label{sec:construction}}

\begin{figure}[h!]
	\centering
		\includegraphics[width=0.45\textwidth]{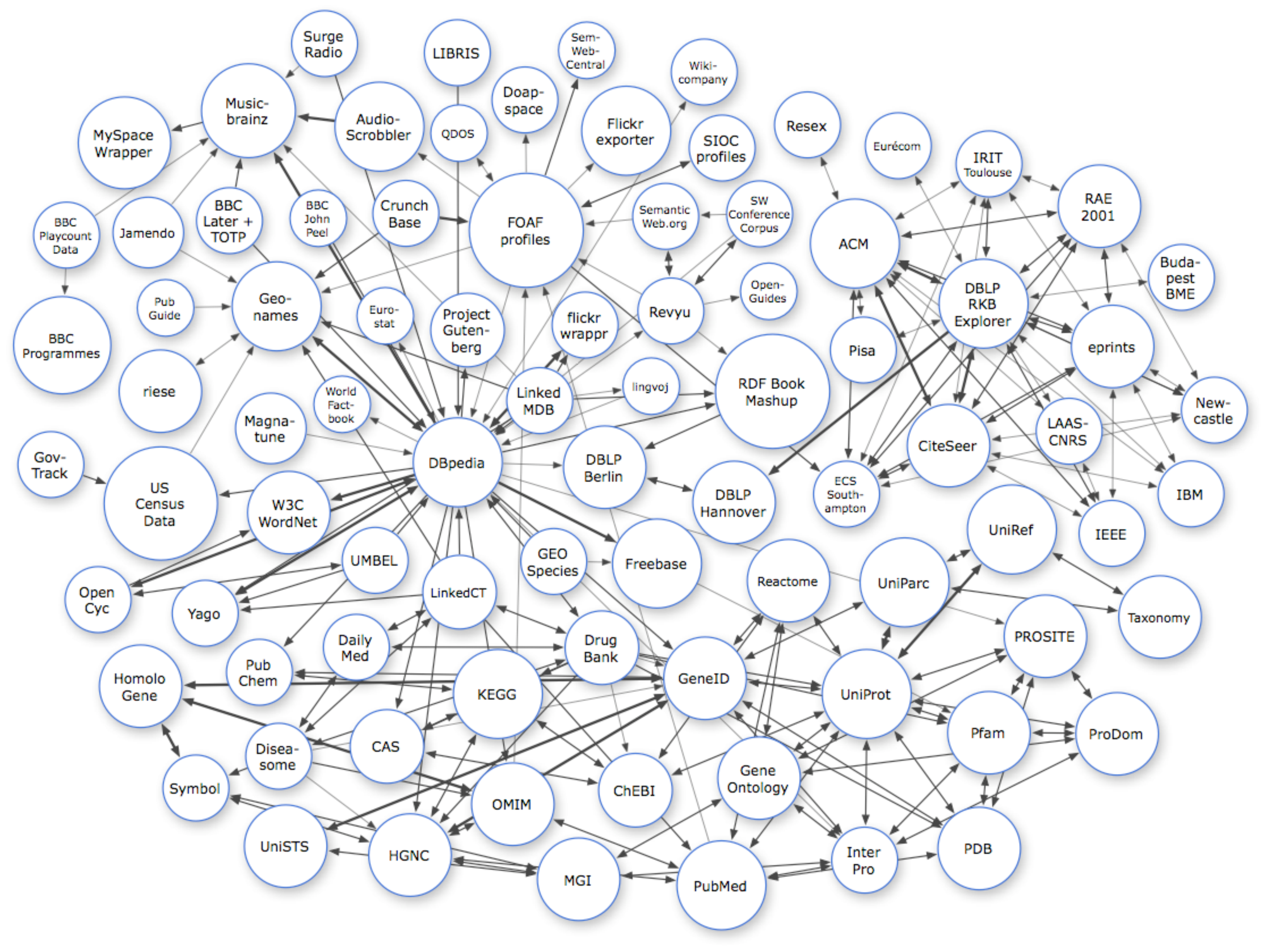}
	\caption{The Linked Data cloud visualization as provided by the Linked Data community. This version is dated February 27, 2009. The author was not responsible for the creation of this visualization. This is only provided in order to better elucidate the means by which the Linked Data cloud graph was created.\label{fig:lod-cloud}}
\end{figure}

The current Linked Data cloud visualization was published by Chris Bizer on February 27, 2009. This visualization is provided in Figure \ref{fig:lod-cloud}.
%%%
%%%
The Linked Data cloud visualization represents various data sets as vertices (i.e.~nodes) and their interlinking relationships as directed unlabeled edges (i.e.~links). Moreover, it is assumed that vertex size denotes the number of triples in the data set and edge thickness denotes the extent to which one data set interlinks with another. Data set $A$ links to data set $B$ if data set $B$ has a URI that is maintained (according to namespace) by data set $A$. In this way, by resolving a data set $B$ URI within data set $A$, the man or machine is able to traverse to data set $B$ from $A$. 

A manual process was undertaken to turn the Linked Data cloud visualization into a Linked Data cloud graph denoted $G = (V, E)$, where $V$ is the set of vertices (i.e.~data sets), $E$ is the set of unlabeled edges (i.e~data set links), and $E \subseteq (V \times V)$. The link weights and the node sizes in the original visualization were ignored. A new visualization of the manually generated Linked Data cloud graph is represented in Figure \ref{fig:lod-graph}. The properties of this visualization are discussed throughout the remainder of this article.

\section{Standard Graph Statistics\label{sec:statistics}}

Given the constructed Linked Data cloud graph visualized in Figure \ref{fig:lod-graph}, it is possible to calculate various graph statistics. A collection of standard graph statistics are provided in Table \ref{tab:graphstats}.
%%%
\begin{table}[h!]
%\begin{footnotesize}
\begin{tabular}{lc}
\hline
statistic	&	statistic value \\
\hline
number of vertices & $86$ \\
number of edges & $274$ \\
weakly connected & true \\
strongly connected & false \\
diameter & $10$ \\
average path length & $3.916$ \\
\hline
\end{tabular}
\caption{\label{tab:graphstats} A collection of standard graph statistics for the Linked Data cloud graph represented in Figure \ref{fig:lod-graph}.}
%\end{footnotesize}
\end{table}

\subsection{Strongly Connected Components}

The Linked Data graph is not strongly connected. This means that there does not exist a path from every data set to every other data set. Therefore, a walk along the graph can lead to an ``island" of data sets that can not be returned from. The number of strongly connected components is $31$ with $26$ of those components only maintaining a single data set (that is, they are either the source of a path or the sink of a path). The size of the remaining strongly connected components is $37$, $15$, $4$, $2$, and $2$. The largest component (with size of $37$) is the ``DBpedia component". The second largest (with size of $15$) is the ``DBLP RKB Explorer component".

Given the large diameter and average path length, the Linked Data cloud graph can be seen as a two weakly connected components: the larger DBpedia component and the smaller DBLP RKB Explorer component. However, as will be seen later, other communities in the larger DBpedia component exist such as biological and medical communities.

\subsection{Degree Distributions}

\begin{figure}[ht!]
	\centering
		\includegraphics[width=0.4\textwidth]{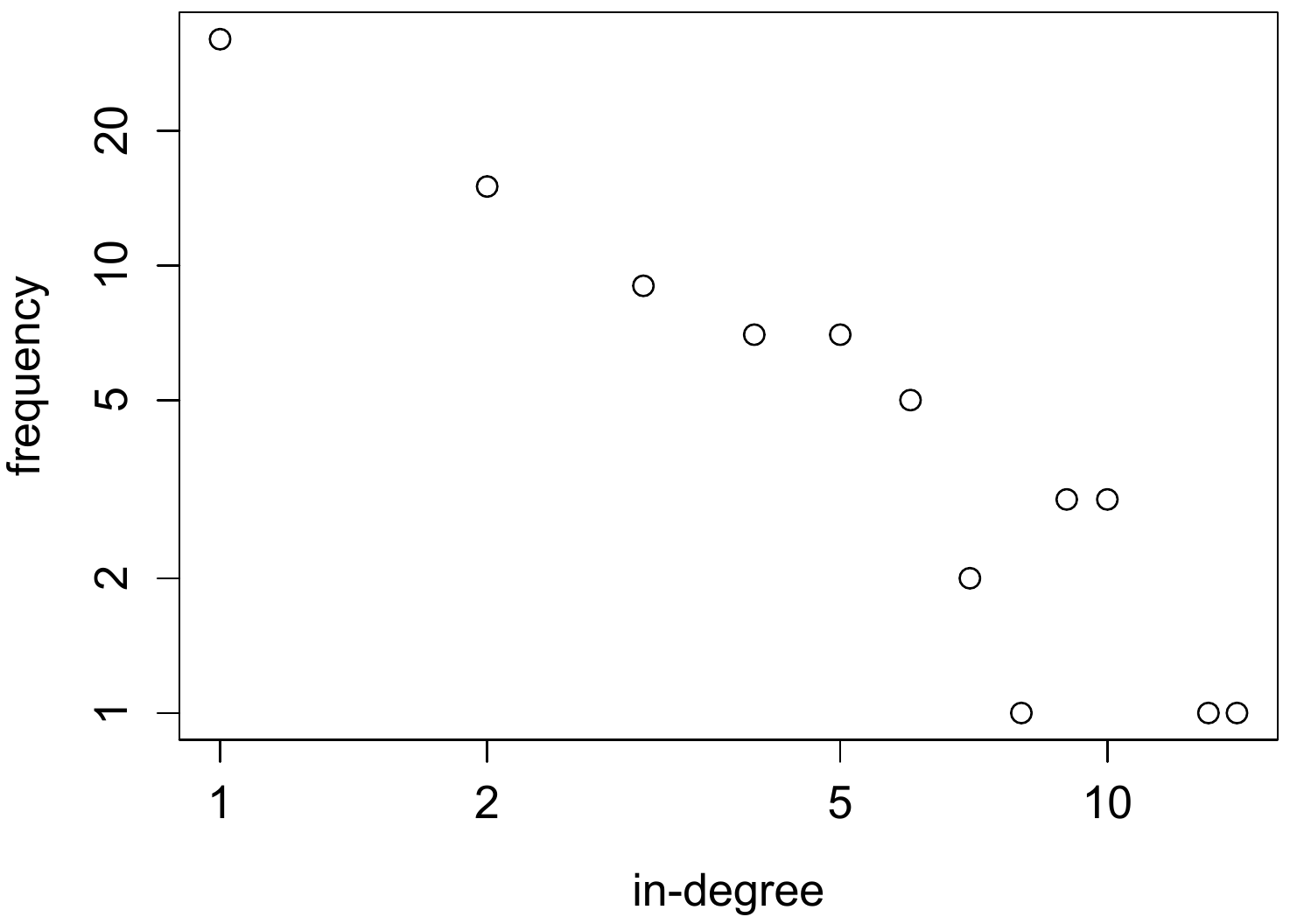}
	\caption{The in-degree distribution of the Linked Data cloud graph on a log-log plot.\label{fig:in-degree}}
\end{figure}

\begin{figure}[ht!]
	\centering
		\includegraphics[width=0.4\textwidth]{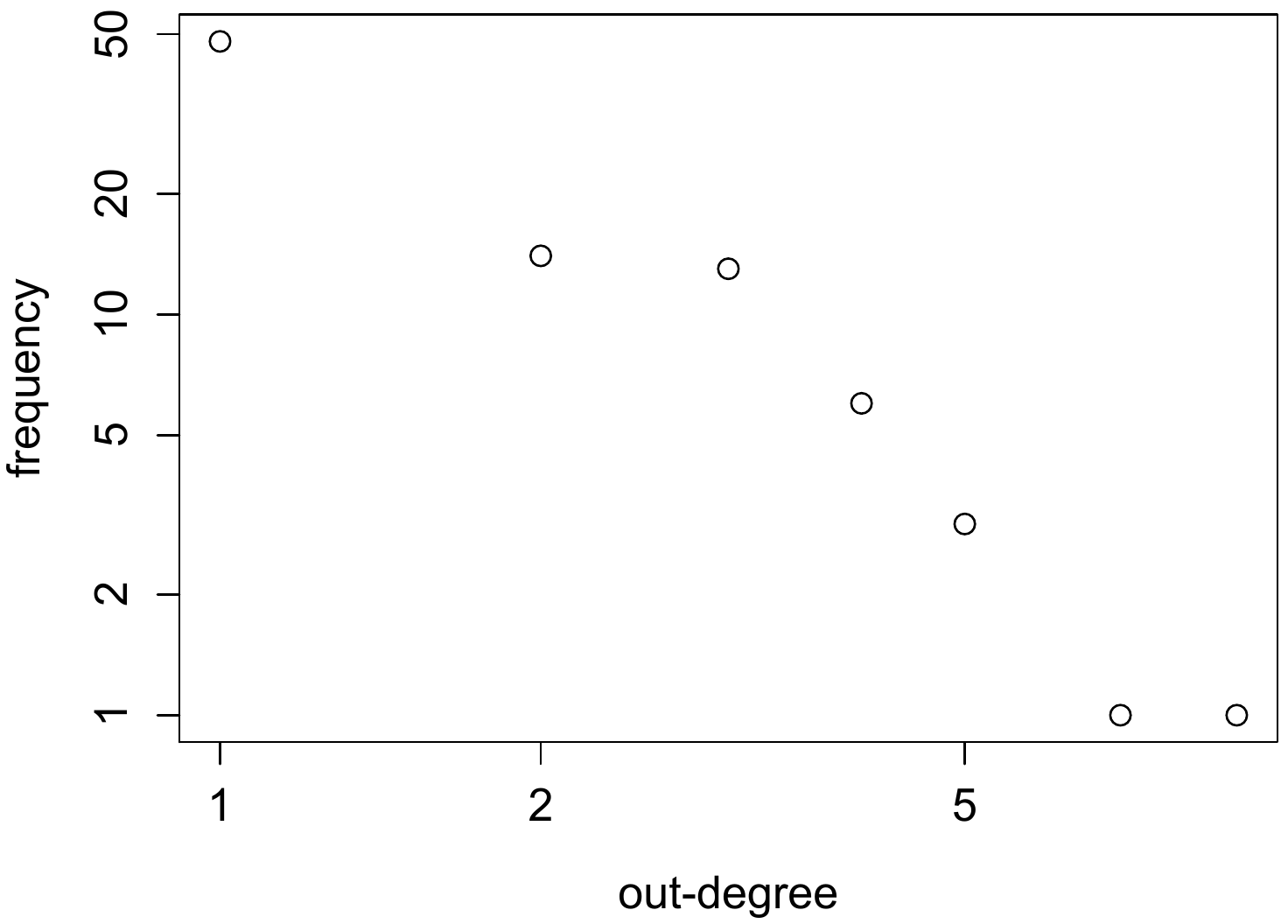}
	\caption{The out-degree distribution of the Linked Data cloud graph on a log-log plot.\label{fig:out-degree}}
\end{figure}

The in- and out-degree distributions of the graph are plotted in Figure \ref{fig:in-degree} and Figure \ref{fig:out-degree} on a log-log plot, respectively. These plots show the number (frequency) of data sets that have a particular in- or out-degree. The top 11 in- and out-degree data sets are presented in Table \ref{tab:in-top} and Table \ref{tab:out-top}, respectively. It is interesting to note that the two leaders (DBpedia and DBLP RKB Explorer) are also the leaders of the two largest strongly connected components identified previously.

\begin{table}[h!]
%\begin{footnotesize}
\begin{tabular}{lc}
\hline
data set	&	in-degree \\
\hline
DBpedia & $14$ \\
DBLP RKB Explorer & $13$ \\
ACM & $10$ \\
GeneID & $10$ \\
GeoNames & $10$ \\
CiteSeer & $9$ \\
ePrints & $9$ \\
UniProt & $9$ \\
ECS Southampton & $8$ \\
FOAF Profiles & $7$ \\
RAE 2001 & $7$ \\
\hline
\end{tabular}
\caption{\label{tab:in-top} The top $11$ Linked Data data sets with the highest in-degree.}
%\end{footnotesize}
\end{table}

\begin{table}[h!]
%\begin{footnotesize}
\begin{tabular}{lc}
\hline
data set	&	out-degree \\
\hline
DBpedia & $17$ \\
DBLP RKB Explorer & $14$ \\
ACM & $10$ \\
CiteSeer & $9$ \\
EPrints & $9$ \\
GeneID & $8$ \\
UniProt & $8$ \\
DrugBank & $7$ \\
ECS Southampton & $7$ \\
FOAF Profiles & $7$ \\
RAE 2001 & $7$ \\
\hline
\end{tabular}
\caption{\label{tab:out-top} The top $11$ Linked Data data sets with the highest out-degree.}
%\end{footnotesize}
\end{table}

While the number of data points is small, a power-law fit is provided according to a distribution that is defined as $p(x) \sim x^{-\alpha}$, where $p(x)$ is the probability of seeing a data set with a degree of $x$. A power-law fit to the total degree distribution (i.e.~ignoring edge directionality) yields an exponent of $\alpha = 1.496$. In other words, the larger the degree, the fewer number of data sets.

\subsection{Degree Correlations}

The correlation between the in- and out-degrees of the vertices yields a Spearman $\rho = 0.6753$ with a significant $p < 9.85^{-13}$. Similarly, the Kendall $\tau = 0.5640$ with a significant $p < 7.27^{-12}$. In other words, data sets that frequently link to other data sets tend to get linked to frequently.

If a graph is degree assortative then vertices with high degree are connected to other vertices with high degree. Likewise, vertices with low degree connect to vertices with low degree. Assortativity is calculated by creating two vectors of length $|E|$. One vector maintains the degree of the vertices at the head of each edge and the other vector maintains the degree of the vertices at the tail of each edge. These two vectors are then correlated. The popular assortative mixing value \cite{newman:assort} is calculated with a Pearson correlation over the two vectors as
%%%
\begin{equation*}
r = \frac{|E| \sum_{i} j_i k_i -  \sum_i j_i \sum_i  k_i}{\sqrt{\left[|E| \sum_i j^2_i - \left(\sum_i j_i\right)^2\right]\left[|E| \sum_i k^2_i -\left(\sum_i k_i\right)^2\right]}},
\end{equation*}
%%%
where $j_i$ is the degree of the vertex on the tail of edge $i$, and $k_i$ is the degree of the vertex on the head of edge $i$. The correlation coefficient $r$ is in $[-1,1]$, where $-1$ represents a fully disassortative graph, $0$ represents an uncorrelated graph, and $1$ represents a fully assortative graph. Given that the degree distribution is non-parametric, a non-parametric assortativity correlation is also provided using both Spearman $\rho$ and Kendall $\tau$. All of these assortativity correlations are presented in Table \ref{tab:assort}, where the only significant values are from the standard Pearson correlation and all the in-degree correlations.
%%%
\begin{table}[h!]
%\begin{footnotesize}
\begin{tabular}{lccc}
\hline
method & in-degree	& out-degree  & total-degree\\
\hline
pearson & -0.1911 (0.0015) & -0.1728 (0.0042) & -0.1868 (0.0019) \\
spearman & -0.1319 (0.0292) & -0.0311(0.6089) & -0.0629 (0.2998) \\
kendall & -0.0933 (0.0346) & -0.0193 (0.6626) & -0.0364 (0.3982) \\
\hline
\end{tabular}
\caption{\label{tab:assort} Various degree assortativity correlations for the Linked Data cloud graph. The first number is the correlation and the second number in parentheses is the $p$-value. A significant $p$-value is less than $0.05$.}
%\end{footnotesize}
\end{table}
%%%
These results demonstrate that Linked Data data sets tend to connect to data sets with differing degrees. That is, for instance, high degree data sets connect to low degree data sets. This is made apparent when looking at DBpedia which has a total-degree of $32$. DBpedia's neighbors in the graph have the following total-degrees: $1$, $1$, $2$, $3$, $3$, $3$, $3$, $3$, $4$, $4$, $4$, $6$, $8$, $9$, $11$, $12$, $12$, and $18$. However, in general, the degree assortativity correlation is weak and for the non-parametric correlations, mostly insignificant.

\section{Structural Analysis\label{sec:structural}}

This section presents an analysis of the community structures that exist within the Linked Data cloud graph. A community is loosely defined as a set of vertices that have a high number of intra-connections and low number of inter-connections. In other words, vertices in the same community tend to link to vertices in the same community as opposed to vertices in other communities. In order to compare the algorithmically determined structural communities to the metadata properties of the vertices that compose those communities, two metadata properties were gathered:
%%%
\begin{enumerate}
	\item a string label denoting the type of content maintained in the data set
	\item an integer value denoting the number of triples contained in the data set. 
\end{enumerate}

The content labels were determined manually. The set of labels used were: biology, business, computer science, general, government, images, library, location, media, medicine, movie, music, reference, and social. Note that many data sets could have been labeled with more than one label. However, only one label was chosen. Moreover, these labels were determined by reviewing the websites of the data sets and not by looking at the structure of the graph. 

The data set triple counts were taken from the ``Linking Open Data on the Semantic Web" web page.\footnote{Linking Open Data on the Semantic Web is available at: \ttt{http://tinyurl.com/5fcmzm}.} Of the $86$ data sets in the Linked Data cloud, only $31$ of those data sets have published triple counts.

\subsection{Labeled Structural Communities}

The graph analysis method for comparing nominal vertex metadata with structural communities as originally presented in \cite{onthe:rodriguez2008} was used to compare the content labels of the data sets to their structural communities. The purpose of this analysis is to determine the semantics of the structural communities. The hypothesis is that structural communities denote shared content. That is, data sets in the same structural community maintain the same type of content data (e.g.~biology, medicine, computer science, etc.).

A contingency table was created that denotes the number of vertices that have a particular content label and are in a particular structural community. An example contingency table that has community values that were determined using the leading eigenvector community detection algorithm \cite{newman-eigen} is presented in Table \ref{tab:conting}. 
%%%
\begin{table}[h!]
\label{tab:chisquare}
%\begin{footnotesize}
\begin{tabular}{ccccccccccc}
\hline  
  \begin{scriptsize} content/community \end{scriptsize}&                0  &1 & 2 & 3 & 4 & 5&  6&  7&  8 & 9\\ 
 \hline
  biology  &        2 & 0 & 4 & 1 & 0 & 0 & 0&  0 & 3 &10\\ 
  business &        1 & 0 & 0 & 0 & 0 & 1 & 0 & 0 & 0 & 0\\ 
  computer science  & 1& 12 & 0 & 0 & 0 & 0 & 0 & 2 & 0&  0\\ 
  general       &   4 & 3 & 0 & 0 & 0 & 0 & 0 & 1 & 0 & 0\\ 
  government  &     3 & 0 & 0 & 0 & 0 & 2 & 0 & 0 & 0 & 0\\
  images        &   1 & 0 & 0 & 0 & 0 & 0 & 0 & 1 & 0 & 0\\ 
  library        &  2  &0&  0&  0&  0 & 1 & 0 & 1 & 0 & 0\\ 
  location     &    0 & 0 & 0 & 0 & 0 & 1 & 0 & 1 & 0 & 0\\ 
  media         &   0 & 0 & 0 & 0 & 0 & 0 & 1 & 0 & 0 & 0\\ 
  medicine   &       0 & 1&  0 & 4 & 0 & 0 & 0 & 0 & 1 & 1\\ 
  movie     &       1 & 0 & 0 & 0 & 0 & 0 & 0 & 0 & 0&  0\\ 
  music    &        5 & 0 & 0&  0&  0 & 1&  1 & 1 & 0 & 0\\ 
  reference  &      2 &  0 & 1 & 1 & 0 & 0 & 0 & 0 & 0 &  0\\ 
  social  &         0  & 0 & 0 & 0 & 1 & 0 & 0 & 5 & 0 & 1\\ 
  \hline
\end{tabular}
\caption{\label{tab:conting} An example contingency table that denotes how many data sets have a particular content label and structural community. For this example, the structural communities were determined using the leading eigenvector community detection algorithm.}
%\end{footnotesize}
\end{table}
%%%
The contingency table is subjected to a $\chi^2$ analysis in order to determine if the manually generated content labels are statistically related to the algorithmically determined structural communities. Four community detection algorithms (and thus, four individual contingency tables) were used for this analysis and the $\chi^2$ $p$-values are presented in Table \ref{tab:chisquare}. 

\begin{table}[h!]
\label{tab:chisquare}
%\begin{footnotesize}
\begin{tabular}{ll}
\hline
community algorithm & $\chi^2$ $p$-value \\
\hline
Leading Eigenvector & $6.6^{-12}$ \\
WalkTrap & $2.2^{-16}$ \\
Edge Betweenness  & $0.0323$ \\
Spinglass & $2.4^{-16}$\\
\hline
\end{tabular}
\caption{\label{tab:chisquare} The $p$-values for four $\chi^2$ tests using four structural community detection algorithms: leading eigenvector \cite{newman-eigen}, walktrap \cite{latapy}, edge betweenness \cite{girvan-2002}, and spinglass  \cite{spinglass:reichardt2006}.}
%\end{footnotesize}
\end{table}

\begin{figure*}[h!]
	\centering
		\includegraphics[width=0.9\textwidth]{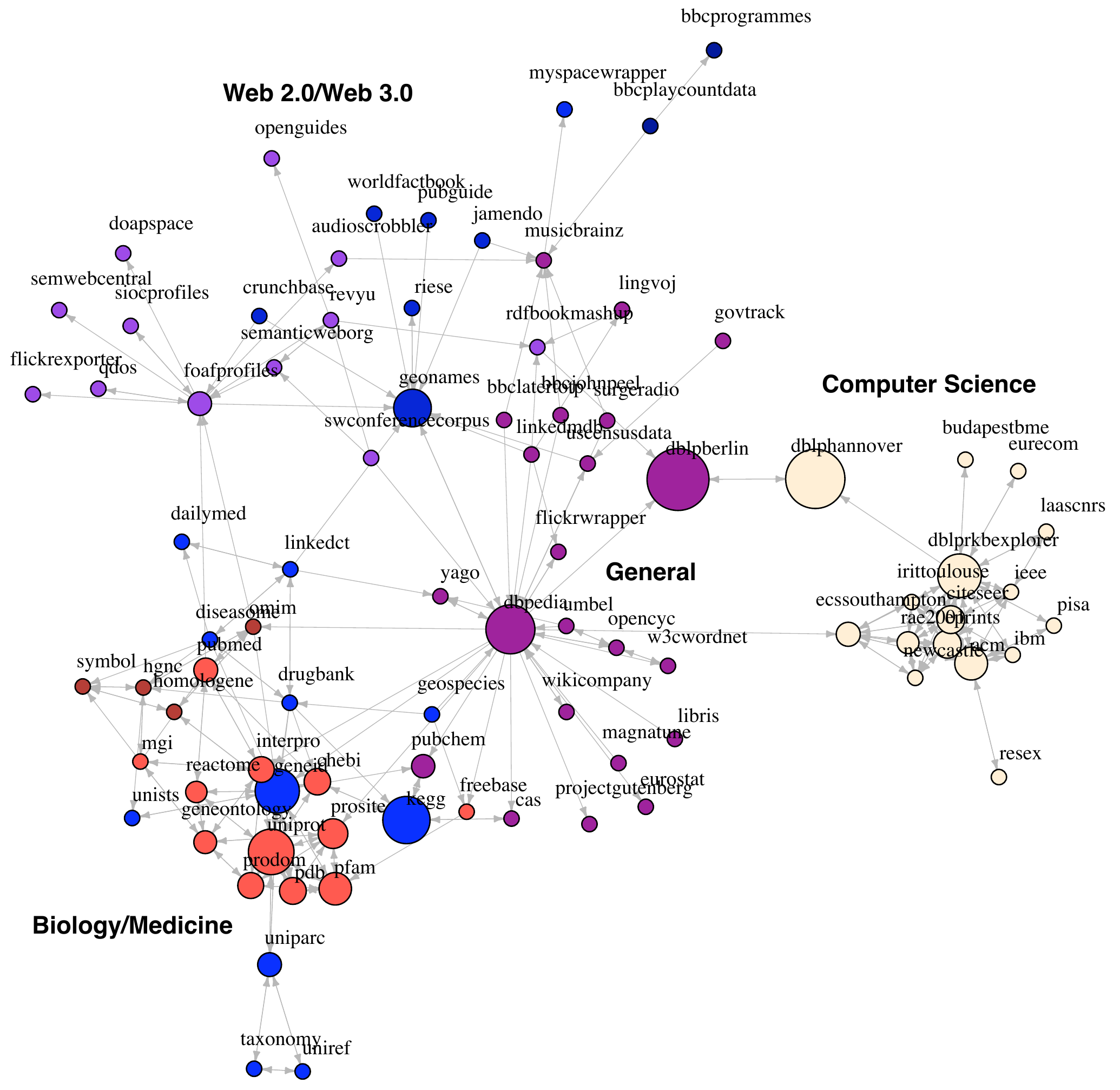}
	\caption{A graph representation of the March 2009 Linked Data Cloud. Each vertex denotes a Linked Data data set. Each edge denotes whether one data set makes reference to another. The size of the vertices are determined by their PageRank centrality according to a $\delta = 0.85$ \cite{page98pagerank}. The vertex colors denote the structural communities as identified by the leading eigenvector community detection algorithm \cite{newman-eigen}. Finally, the Fruchterman-Reingold layout algorithm was used to visually render this representation \cite{layout:fruchter1991}.\label{fig:lod-graph}}
\end{figure*}

The analysis demonstrates that data sets that maintain similar content tend to exist in the same structural areas of the graph. This is made salient by a qualitative analysis of various subsets of the graph (see Figure \ref{fig:lod-graph} where the vertex colors denote their structural community). Moreover, this makes sense intuitively. Data sets that share the same content labels are more than likely to reference to the same resources. For example, it is true that medical data sets tend to be connected to other medical data sets and not to music data sets. Table \ref{tab:comstruct} provides a review of 15 randomly chosen Linked Data data sets, their structural community values according to the leading eigenvector community detection algorithm, and their manually determined content labels.
%%%
\begin{table}[h!]
%\begin{footnotesize}
\begin{tabular}{lccc}
\hline
data set & community & content label \\
\hline
SurgeRadio & 0 & music \\
MusicBrainz & 0 & music \\
DBpedia & 0 & general \\
Riese & 5 & government \\
LinkedCT & 3 & medicine \\
World Fact Book & 5 & government \\
OpenCyc & 0 & general \\
Yago & 0 & general \\
DrugBank & 3 & medicine \\
DailyMed & 3 & medicine \\
UniParc & 2 & biology \\
Reactome & 9 & biology \\
ACM & 1 & computer science \\
CiteSeer & 1 &computer science \\
IEEE & 1 & computer science \\
\hline
\end{tabular}
\caption{\label{tab:comstruct}A sample of 15 Linked Data data sets, their leading eigenvector structual community value, and their manually determined content label.}
%\end{footnotesize}
\end{table}

\subsection{Data Set Triple Counts}

Of the $86$ data sets in the Linked Data cloud, only $31$ of those data sets have triple counts that were published on the ``Linking Open Data on the Semantic Web" web page.  Given the statistically significant, positive correlation between the in-degree and out-degree of the vertices, it is hypothesized that those data sets that are more central in the graph will have a larger triple count. The centrality of all $86$ vertices was determined using the PageRank centrality algorithm with a $\delta = 0.85$ \cite{page98pagerank}. For those $31$ data sets that have triple counts, their triple count value was correlated with their PageRank centrality value. The Spearman $\rho = 0.6274$ with a significant $p < 0.00016$. Similarly, the Kendall $\tau = 0.4566$ with a significant $p < 0.00039$. Thus, those data sets that have the most RDF triples tend to be centrally located in the Linked Data cloud.

Finally, an assortative mixing calculation over data set triple counts was performed. Given that only $31$ data sets have triple count values, a $31$ vertex subgraph was created. This $31$ vertex graph has $56$ edges. These $56$ edges were used to determine the assortative triple count correlation. Thus, two vectors of length $56$ were created where one vector maintained the triple count of the data sets on the head of each edge and the other vector maintained the triple count of the data sets on the tail of each edge. Table \ref{tab:assortsize} provides three assortativity correlations. Note that the triple count data distribution is non-parametric. From these results, only the non-parametric Kendall correlation is statistically significant with a correlation that demonstrates that the data sets are loosely disassortative according. This means that small data sets tend to connect to large data sets and large data sets tend to connect to small data sets. Again, this correlation is relatively weak.
%%%
\begin{table}[h!]
%\begin{footnotesize}
\begin{tabular}{lc}
\hline
method & size assortativity \\
\hline
pearson & 0.0682 (0.3230) \\
spearman & -0.2546 (0.0559) \\
kendall & -0.2064 (0.0302)  \\
\hline
\end{tabular}
\caption{\label{tab:assortsize} Data set triple count assortativity correlations for the Linked Data cloud graph. Given that only $31$ data sets have published triple counts, these assortativity values are determined according to this $31$ data set subgraph. The first number is the correlation and the second number in parentheses is the $p$-value. A significant $p$-value is less than $0.05$.}
%\end{footnotesize}
\end{table}

\subsection{Data Set Centrality}

The PageRank centrality (with $\delta = 0.85$) of each of the $86$ data sets in the Linked Data cloud graph was calculated. Table \ref{tab:pagerank} provides the top $15$ central data sets. From this analysis, and assuming that centrality denotes ``importance", it appears that content in computer science and biology are of major import to the current instantiation of the Linked Data cloud.

\begin{table}[h!]
%\begin{footnotesize}
\begin{tabular}{lcc}
\hline
data set & page rank & content label \\
\hline
DBLP Berlin & 0.0484  & computer science \\
DBLP Hannover &  0.0464 & computer science \\
DBpedia & 0.0384  & general  \\      
KEGG  &  0.0370  & biology \\       
UniProt  & 0.0357 & biology  \\      
GeneID   & 0.0346 & biology  \\      
DBLP RKB Explorer & 0.0341 & computer science \\
GeoNames  &  0.0294 & location  \\      
ACM  & 0.0257  & computer science \\
Pfam & 0.0254  & biology  \\      
Prosite &  0.0233 &  biology  \\      
ePrints  & 0.0218 & computer science \\       
CiteSeer  &  0.0218  & computer science \\       
PDB & 0.0209  & biology \\     
\hline
\end{tabular}
\caption{\label{tab:pagerank} The top 15 PageRank central data sets in the Linked Data cloud graph.}
%\end{footnotesize}
\end{table}

\section{Conclusion}

The Linked Data initiative is focused on unifying RDF data sets into a single global data set that can be utilized by both man and machine. This initiative is providing a fundamental shift in the way in which data is maintained, exposed, and interrelated. This shift is both technologically and culturally different from the relational database paradigm. For one, the address space of the Web of Data is the URI address space, which is inherently distributed and infinite. Second, the graph data structure is becoming a more accepted, flexible representational medium and as such, may soon displace the linked table data structure of the relational database model. Finally, with respects to culture, the Web of Data maintains publicly available interrelated data. In the relational database world, rarely are database ports made publicly available for harvesting and rarely are relational schemas published for reuse. The Semantic Web, the Linked Data community, and the Web of Data are truly emerging as a radical rethinking of the way in which data is managed and distributed in the modern world.

\end{document}